# Wet-spinnability and crosslinked fibre properties of two collagen polypeptides with varied molecular weight

Giuseppe Tronci,[1,2*] Ramya Sri Kanuparti,[1,2] M. Tarik Arafat,[1,2] Jie Yin,[1,2] David J. Wood,[2] Stephen J. Russell[1]

[1] Nonwovens Research Group, School of Design, University of Leeds, UK

[2] Biomaterials and Tissue Engineering Research Group, School of Dentistry, St. James's University Hospital, University of Leeds, UK

**Abstract**

The formation of naturally-derived materials with wet stable fibrous architectures is paramount in order to mimic the features of tissues at the molecular and microscopic scale. Here, we investigated the formation of wet-spun fibres based on collagen-derived polypeptides with comparable chemical composition and varied molecular weight. Gelatin and hydrolysed fish collagen (HFC) were selected as widely-available linear amino-acidic chains of high and low molecular weight, respectively, and functionalised in the wet-spun fibre state in order to preserve the material geometry in physiological conditions. Wet-spun fibre diameter and morphology were dramatically affected depending on the polypeptide molecular weight, wet-spinning solvent (i.e. 2,2,2-Trifluoroethanol and dimethyl sulfoxide) and coagulating medium (i.e. acetone and ethanol), resulting in either bulky or porous internal geometry. Dry-state tensile moduli were significantly enhanced in gelatin and HFC samples following covalent crosslinking with activated 1,3-phenylenediacetic acid (Ph) (*E*: 726 ± 43 – 844 ± 85 MPa), compared to samples crosslinked via intramolecular carbodiimide-mediated condensation reaction (*E*: 588 ± 38 MPa). Resulting fibres displayed a dry diameter in the range of 238±18–355±28 µm and proved to be mechanically-stable ($E$: 230 kPa) following equilibration with PBS, whilst a nearly-complete degradation was observed after 5-day incubation in physiological conditions.

---

[*] Corresponding author: Giuseppe Tronci, Level 7 Wellcome Trust Brenner Building, St. James's University Hospital, Leeds LS9 7TF, UK (Email: g.tronci@leeds.ac.uk)

## 1. Introduction

Polypeptides are essential components in various biological materials and play a major role in structural stabilisation, energy conversion and tissue remodelling [1]. Polypeptides derived from the organic matrix of tissues, such as gelatin or hydrolysed collagen are widely available, biodegradable and contain cell-binding sequences, making them appealing building blocks for the design of healthcare products found in applications such as chronic wound care, guided bone regeneration (GBR), and stratified medical devices. The integration of such building blocks into defined fibrous architecture is therefore of high technological and scientific interest in order to mimic tissue organisation at different length scales and accomplish material systems with unexpected properties and functionalities.

Polypeptide-based fibrous architectures can typically be prepared via solution-based spinning techniques, whereby two main strategies can be pursued: (i) direct formation of nano-/submicron webs by electrospinning [2]; (ii) fibre spinning, via methods such as dry- [3], gel- [4], or wet-spinning [5, 6, 7], potentially combined with post-drawing [8] or electrophoretic compaction [9], whereby fibres are obtained and used to make customised fabrics. Both strategies have been successfully applied to gelatin as collagen-derived polypeptide, resulting in spirally-electrospun webs [10] as well as bulky wet-spun fibres [5] with porous morphology and aligned polypeptide chains. Here, continuous gelatin filaments were successfully formed via a ternary system of water, isopropanol and plasticiser, whereby superior and tunable dry-state tensile moduli (up to nearly 4 GPa) were obtained depending on the wet-spinning mixture formulation and plasticiser content.

Electrospinning has received a great deal of attention because it is simple to operate and inexpensive, and can process a range of polymers [11]. It also enables the formation of cell-laden webs via co-spinning of synthetic polymers with cell suspensions [12], an approach currently refereed as "cell electrospinning" [13]. Despite largely applicable to synthetic polymers, however, electrospinning of polypeptides offers restricted possibilities with regard to architecture customisation, three-dimensional geometry and morphology preservation in physiological environments [14]. Other than electrospinning, classical fibre spinning methods, such as dry- or wet-spinning, can lead to the formation of defined fibrous building blocks, depending on the type of wet-spinning solvent and coagulating medium [15, 16, 17]. Depending on the polypeptide chemical treatment and orientation at the molecular scale, resulting fibres can display good wet stability in aqueous environment offering possibilities for the design of textile or nonwoven architectures suited to clinical use. Besides fibre formation, reliable synthetic methods should be applied to the resulting fibrous system so that defined polypeptide-based network architectures can be introduced at the molecular scale, thereby ensuring adjustable mechanical properties and wet-stable morphology [18].

In order to prepare polypeptide fibres with adequate mechanical properties to enable them to retain their structure and function in applicable clinical and biological environments, we investigated the wet-spinnability of gelatin and HFC, and characterised resultant fibre properties in the as-spun and crosslinked states. The effect of the wet-spinning solvent, coagulating medium and polypeptide molecular weight on corresponding fibre morphology was addressed aiming to accomplish homogeneous fibres with controlled morphology. Varied polypeptide-based covalent networks were synthesised in the fibrous state via (i) reaction with activated 1,3-phenylenediacetic acid (Ph) in the presence of carbodiimide (potentially resulting in either intra- or inter-molecular crosslinks); (ii) via reaction with activated Ph only; (iii) via carbodiimide-mediated condensation, leading to the formation of intramolecular covalent crosslinks. Reacted fibres were investigated with regard to crosslink density, tensile and swelling properties as well as hydrolytic degradability in order to establish the governing structure-property relationships of the fibrous system.

**2. Materials and methods**

**2.1 Materials**

Type A gelatin from porcine skin (175 g Bloom), 2,2,2-Trifluoroethanol (TFE), Dimethyl Sulfoxide (DMSO), acetone, ethanol and 2,4,6-trinitrobenzenesulfonic acid (TNBS) were purchased from Sigma Aldrich. 1,3-phenylenediacetic acid (Ph), N-hydroxysuccinimide (NHS) and 1-ethyl-3-(3-dimehylaminopropyl) carbodiimide hydrochloride (EDC) were purchased from Alfa Aesar. HFC was kindly provided by Nitta Gelatin India Limited.

**2.2 Sodium dodecyl sulphate-polyacrylamide gel electrophoresis (SDS-page)**

Gelatin and HFC were dissolved in SDS sample buffer (160 mM Tris–HCl, pH 6.8, 2% SDS, 26% glycerol, 0.1% bromophenol blue) at 1 %wt./vol. concentration and heated for 2 min at 90 °C. 30 mL of each sample solution were loaded onto 4% stacking gel wells and separated on 15% resolving gels (200 V, 45 min, room temperature). Protein bands were visualized after 60 min staining (0.1 wt.-% Comassie Blue, 12.5 vol.-% trichloroacetic acid) and 60 min treatment in water. Two gel lanes were employed for each polypeptide and imaged with ChemiDoc$^{TM}$ MP System using ImageLab software version 4.1 (Bio-Rad). Selected bands (band intensity: $3 \cdot 10^5 - 13 \cdot 10^6$) were analysed with regard to their molecular weights, so that the number-average molecular weight ($M_n$) was determined by averaging obtained molecular weight values.

**2.3 Viscosity measurements on wet-spinning solutions**

Wet-spinning solutions (3–40 %wt./vol. protein) were prepared in either DMSO or TFE via stirring at 37 °C overnight. The dynamic viscosity of resulting solutions was measured using a DV-E Viscometer (Brookfield Viscometers Ltd., Harlow, UK). The solution was filled into a beaker making sure that no air bubbles were present and the solution temperature was constant.

**2.4 Formation of wet-spun fibres**

Resulting solutions were transferred into a 10 ml syringe having 14.5 mm internal diameter equipped with Ø 0.8 μm needle, which acted as single nozzle spinneret. The collagen solutions were then extruded from the nozzle tip via a syringe pump at a dispensing rate of 12 ml·hr$^{-1}$ with the nozzle tip submerged in a coagulation bath containing 1 litre of either ethanol or acetone at room temperature. Resulting wet-spun fibres were then removed from the coagulating medium and dried separately at room temperature. A nearly quantitative yield (> 99 wt.-%) of fibre formation was measured with respect to the polypeptide weight used in the wet-spinning solution.

**2.5 Scanning Electron Microscopy (SEM)**

Dry samples were fixed on carbon stubs and gold sputtered under argon prior to SEM. The structure and morphology of wet-spun fibres were examined using SEM (JEOL SM-35) at different magnifications. The diameter of each fibre group was quantified from SEM images in six different locations, so that average and standard deviation were used for data representation. The internal pore size in each fibre was quantified in the same manner whereby fifteen measurements were taken per each sample group.

**2.6 Crosslinking of wet-spun fibres**

Three different crosslinking methods were applied to wet-spun fibres at room temperature. In the first instance, samples were incubated with NHS-activated Ph in the presence of EDC-NHS as follows: 0.074g of 1,3- Phenylenediacetic acid (Ph) was dissolved in 100ml ethanol with magnetic stirring at 40°C till a clear solution was obtained. Resulting Ph solution was cooled down to room temperature and 4.5 mmoles of both 1-Ethyl-3-(3-dimethylaminopropyl) carbodiimide hydrochloride (EDC) and N-hydroxysuccinimide (NHS) were added. The solution was stirred for 30 min in order to activate Ph carboxylic acid terminations, following which time 0.06 g of fibres was added into the solution for the crosslinking reaction to occur. Alternatively, samples were reacted with NHS-activated Ph following the above procedure with the only difference that an equimolar amount of β-mercaptoethanol (βME) with respect to EDC/NHS was added following Ph activation and prior to the crosslinking reaction, in order to deactivate EDC/NHS. Ultimately, samples were crosslinked via state-of-the art EDC/NHS-mediated condensation reaction (EN), whereby 0.06 g fibre was incubated in 100 ml ethanol solution containing 4.5 mmoles of both EDC/NHS. All reacted samples were collected, rinsed thoroughly with ethanol and air-dried.

**2.7 Swelling tests**

Swelling tests of the fibres were performed by incubating dry samples in 50 ml PBS for 24 hr at 37 °C. PBS-equilibrated samples were paper blotted and weighed. The weight based swelling ratio (*SR*) was calculated as follows:

$$SR = \frac{w_s - w_d}{w_d} \times 100 \qquad \text{(Equation 1)}$$

Where, $w_s$ and $w_d$ are PBS-equilibrated and dry sample weight, respectively. Three replicas were used for each sample group, and results were reported as mean ± standard deviation.

## 2.8 Tensile tests

Tensile modulus and tensile strength of both dry and PBS-equilibrated crosslinked samples were measured using a Zwick Roell Z010 testing system with 10 N load cell at a rate of 0.03 mm·s$^{-1}$. The testing was carried out in a controlled environment with a room temperature of 18 °C and relative humidity of 38%. The gauge length was 5 mm. Five individual fibre samples were tested from each group and measurements were reported as mean ± standard deviation.

## 2.9 TNBS colorimetric assay

TNBS assays were conducted to quantify the degree of crosslinking in reacted fibres, as reported previously [19]. The content of free amino groups and degree of crosslinking (*C*) were calculated as follows:

$$\frac{moles\ (Lys)}{g\ (polypeptide)} = \frac{2 \times 0.02 \times Abs(346\ nm)}{1.46 \times 10^4 \times x \times b} \qquad \text{(Equation 2)}$$

$$C = \left(1 - \frac{moles(Lys)_{crosslinked}}{moles(Lys)_{Polypeptide}}\right) \times 100 \qquad \text{(Equation 3)}$$

where *Abs(346 nm)* is the absorbance value at 346 nm, *0.02* is the volume of sample solution (in litres), *2* is the dilution factor, $1.46 \times 10^4$ is the molar absorption coefficient for 2,4,6-trinitrophenyl lysine (in M$^{-1}$ · cm$^{-1}$), *b* is the cell path length (1 cm) and *x* is the sample weight. Here *mol(Lys)$_{Polypeptide}$* and *mol(Lys)$_{Crosslinked}$* represent the total molar content of free amino groups in native and crosslinked polypeptide fibre, respectively. The nomenclature (*Lys*) is hereby used to recognise that lysines make the highest contribution to the molar content of polypeptide free amino groups, although contributions from hydroxylysines and amino termini are also taken into account.

## 2.10 Degradability study

Dry samples were incubated in PBS (37 °C, pH 7.4) for five days, following which time they were collected, rinsed with distilled water and air-dried. The mass loss ($M_L$) was calculated according to Equation 4:

$$M_L = \frac{m_d - m_t}{m_d} \times 100 \qquad \text{(Equation 4)}$$

whereby $m_d$ and $m_t$ correspond to the dry sample mass before and after PBS incubation, respectively. Three replicas were used for each sample group so that $M_R$ was described as average ± standard deviation.

## 3. Results and discussion

Wet-spun biomimetic fibres were obtained via wet-spinning of gelatin and HFC, as collagen-derived polypeptides of high and low molecular weight, respectively. Polypeptide solutions were prepared in either DMSO or TFE and wet-spun via a syringe pump against common organic non-solvents for collagen (Fig. 1). In the following, the formation of covalently-crosslinked wet-spun polypeptides are described as potential building blocks for the design of biomimetic fabrics, with the aim of investigating how selected polypeptides, wet-spinning conditions and crosslinking methods affect resulting fibre morphology and hydrated macroscopic properties, respectively.

Sample nomenclature in this work is as follows: 'XYZ' indicates a wet-spun sample, whereby 'X' identifies the type of protein used during fibre formation, i.e. 'G' (gelatin) or 'H' (HFC); 'Y' states the solvent used to prepare the wet-spinning solution, i.e. 'T' (TFE) or 'D' (DMSO); 'Z' relates to the coagulating medium, i.e. 'A' (acetone) or 'E' (ethanol). 'XYZ-AA' indicates a crosslinked fibre, whereby 'AA' identifies fibres reacted with NHS-activated Ph in the presence of EDC/NHS (Ph*), NHS-activated Ph (Ph), or via EDC/NHS-mediated condensation (EN).

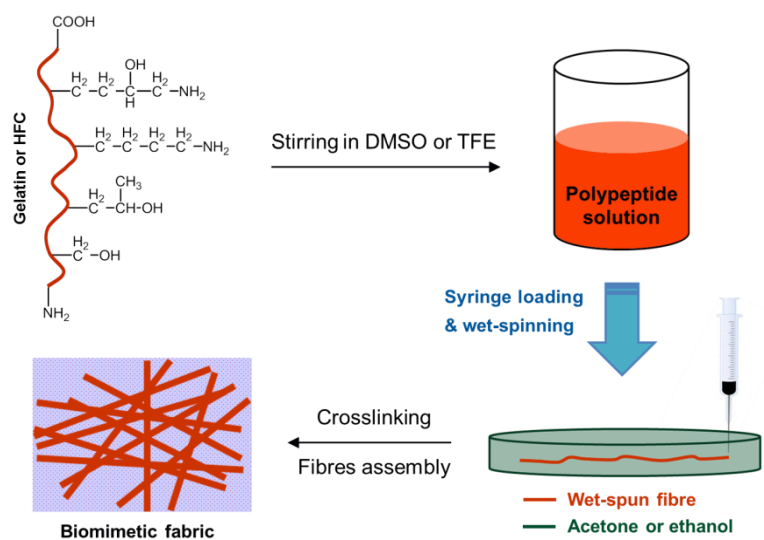

**Fig. 1.** Formation of covalently-crosslinked wet-spun fibres based on collagen-derived polypeptides of varied molecular weight. Both gelatin and HFC solutions were coagulated in specific non-solvents via a syringe pump. Resulting fibres were crosslinked with activated Ph, showing preserved hydrated morphology. These wet-stable fibres could be used as building block for the design of biomimetic fabric.

### 3.1. Characterisation of wet-spinning solutions

Wet-spinning is typically accomplished via the dissolution of a natural or synthetic polymer in a suitable solvent, followed by extrusion and coagulation of the obtained solution in a polymer non-solvent. Since the polymer is non-soluble in the coagulating medium, polymer phase separation occurs, so that fibres are formed by adjusting the solution ejection rate. The concentration of the wet-spinning solution and the type of wet-spinning solvent are expected to play a major role on the morphology, molecular alignment and mechanical properties of resulting fibres [5, 6, 7, 20]. Here, viscosity measurements on varied solutions of both gelatin and HFC were combined with SDS-page analysis in order to investigate how the physical properties of the solution and polypeptide molecular parameters affected the wet-spinnability of the system. Fibres could not be formed with concentrations lower than 10 %w/v, whilst 15-40 %w/v polypeptide solutions were found to be compatible with wet-spinning of either gelatin (μ: 0.07-15 Pa·s) or HFC (μ: 0.01-0.43 Pa·s) (Fig. 2 A), which is in agreement with previous works [3, 23]. DMSO was found to lead to higher solution viscosity compared to TFE when both starting materials were employed, whilst the opposite trend in viscosity was observed when comparing HFC with the gelatin solutions, regardless of the solvent. TFE, as a fluoroalcohol, has been widely used as organic solvent to electrospin biopolymers, including gelatin and collagen, in light of its ability to disrupt inter and intramolecular hydrogen bonds, leading to suppressed chain helicity [21, 22, 23]. Similarly, DMSO is also one of the few organic solvents suitable for gelatin and other polypeptides [4, 24]; however, an increased solvent viscosity was observed with DMSO compared to TFE (Table 1), which likely explains the increased viscosity values of wet-spinning solutions prepared with the former solvent. The solvent-induced variation in solution viscosity is expected to impact on the molecular alignment of protein chains during wet-spinning, due to the inherent variation in polypeptide chain flexibility and axial orientation [25]. Other than the solvent-induced variation in viscosity, gelatin solutions were found to display higher shear stress compared to the HFC solutions (Fig. 2 A).

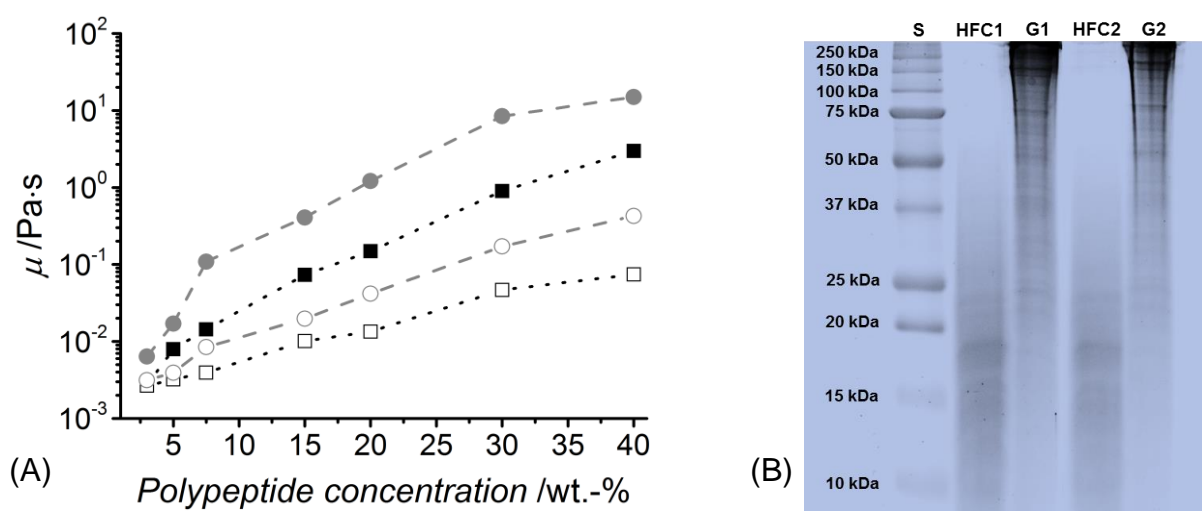

**Fig. 2.** (A): viscosity ($\mu$) of wet-spinning solutions with varied protein concentration in DMSO (--●--: gelatin, --○--: HFC) and TFE (···■···: gelatin, ···□···: HFC). (B): SDS-page analysis of molecular weight standard (S), hydrolysed fish collagen (HFC1-2) and gelatin (G1-2).

In relation to this finding, FTIR spectroscopy confirmed the main bands of collagen, i.e. amide I (at 1650 cm$^{-1}$), II (1550 cm$^{-1}$) and III (at 1240 cm$^{-1}$) bands [26], were present in both polypeptide spectra (Fig. 3, (a) and (b)). On the other hand, SDS-page analysis revealed a distinct molecular weight distribution between gelatin and HFC, whereby bands in the range of 25-250 kDa ($M_n$: 106 kDa) and 10-20 kPa ($M_n$: 17 kDa) were observed, respectively (Fig. 2 B), proving the presence of shorter amino acidic chains in the raw hydrolysed collagen, in agreement with previous reports [27, 28]. The polymer molecular weight is a key parameter in solution-based spinning technologies, because of its influence on the viscosity of the wet-spinning solution as well as the mechanical properties of, and degree of polymer orientation in, resulting spun fibres [29, 30]. In light of the differences in polypeptide size, it was of interest to compare the fibril-forming capability of HFC with respect to gelatin, with the aim of investigating whether the low molecular weight of HFC could still enable the formation of homogeneous fibres with suitable mechanical strength.

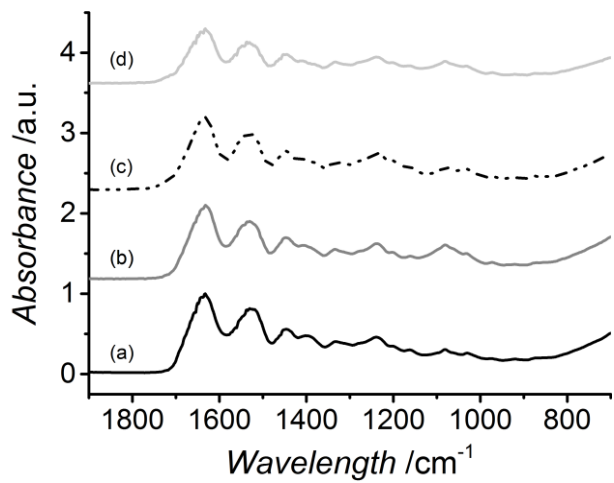

**Fig. 3.** ATR-FTIR spectra of collagen-derived polypeptides in their native ((a): HFC, (b): G) and crosslinked fibre ((c): H30TA-Ph$^*$, (d): G30TA-Ph$^*$) state.

**3.2 Formation of wet-spun fibres**

Fig. 4 depicts the typical fibre morphology of the wet-spun samples. Besides HFC-based fibres obtained with DMSO as the solvent (Fig. 4, F), nearly-homogeneous fibre surfaces were observed via wet-spinning of 15-20 %w/v polypeptide solutions against both acetone (Fig. 4 A, C-D) and ethanol (Fig. 4, B), whilst a spectrum of porous internal morphologies (Fig. 4, C, E) and fibre diameter (Fig. 5) was observed. As expected, HFC solutions in TFE mainly resulted in porous (*P*: 4±1 µm) fibres (Fig. 1 D, E) with decreased diameter (Fig. 5), in light of the low amino acidic chain length, polypeptide concentration and solution viscosity [7]. Other than HFC, gelatin, as the high molecular weight polypeptide, could be wet-spun in both DMSO and TFE, although internal pores were still observed in fibres prepared with the former solvent (Fig. 1 C). The morphology of wet-spun fibres is a characteristic of the dynamics during wet spinning and can be explained in part by considering the diffusional interchange that takes place during the process between the freshly forming filaments containing solvent and the coagulating medium. The coagulating medium diffuses into the filament, while the solvent diffuses out [31]. Therefore, the internal fibre morphology can be adjusted by controlling the diffusion coefficient and rates at which solvent is removed and coagulating medium enters during spinning. In the present study, the morphology of the wet-spun fibres prepared from the DMSO-based polypeptide solutions reveals the formation of an outer dense skin encapsulating a porous substructure [32].

**Table 1.** Solvent polarity parameter ($E_T(30)$), viscosity ($\mu$) and vapour pressure ($e$) of wet-spinning solvents (i.e. DMSO and TFE) and coagulating media (i.e. acetone and ethanol) at 25 °C. [a] $E_T(30)$ values obtained according to ref. [31].

| Solvent | DMSO | TFE | Acetone | Ethanol |
|---|---|---|---|---|
| $E_T(30)$ /kcal·mol$^{-1}$ [a] | 45.1 | 59.8 | 42.2 | 51.9 |
| $\mu$ /Pa·s ($\times 10^{-3}$) | 1.99 | 1.75 | 0.31 | 1.07 |
| $e$ /hPa | 0.6 | 93 | 304 | 78 |

The low gelatin concentration of the starting wet-spinning solution (20 %w/v) and the high viscosity of DMSO as well as the similar solvent polarity [33] between DMSO and acetone (Table 1) likely promotes an instantaneous solvent/non-solvent exchange at the surface of the forming fibre, resulting in a dense surface layer. The formation of such a layer is likely to reduce the diffusion rate of DMSO out of the fibre core, so that inner pores are formed, as has also been observed in spider silk fibres wet-spun from low concentration solutions [7]. The reduced diffusion rate of DMSO out of the fibre core is also likely to explain the increased diameter of resulting fibres (Fig. 5). In contrast to DMSO, less porous, homogeneous and thinner fibres were observed using TFE-based wet-spinning gelatin solutions. This observation is in agreement with the fact that the lower viscosity of TFE (with respect to DMSO) as well as the higher polarity difference between TFE and the non-solvent (compared to DMSO and the non-solvent) (Table 1) result in a decreased diffusion rate, so that diffusion gradients are minimised across the forming fibre and an homogeneous phase-separation occurs. The increased polarity difference between wet-spinning solvent and coagulating medium (Table 1), i.e. TFE and acetone, on the one hand, and TFE and ethanol, on the other hand, is also likely to explain the more regular surface fibre morphology resulting from the former wet-spinning system (Fig. 4, A-B).

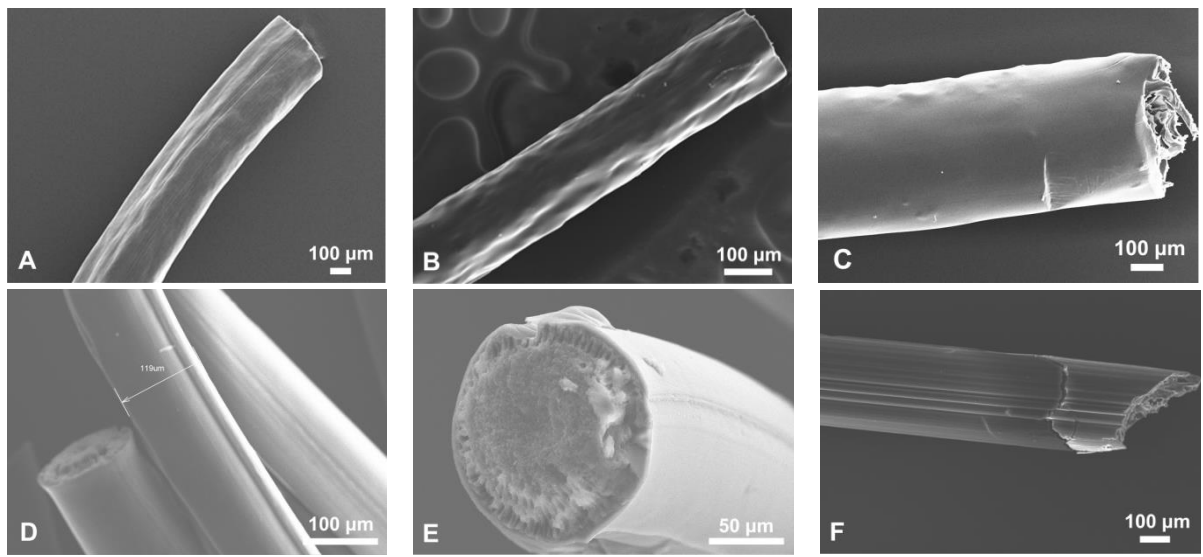

**Fig. 4.** SEM of wet-spun fibres obtained with varied protein, solvent and coagulating medium conditions. (A-B): gelatin fibres obtained from TFE solutions with either acetone (A, G20TA) or ethanol (B, G20TE) as the coagulating medium; (C): gelatin samples G15DA wet-spun from DMSO solutions with acetone as the coagulating medium. (D-E): fibres H15TA resulting from TFE solutions with acetone as the coagulating medium. (F): fibres H20DA wet-spun from DMSO solutions against acetone.

From all the aforementioned observations, it was apparent that a controlled exchange between wet-spinning solvent and coagulating medium was paramount to the formation of dense and homogeneous polypeptide fibres. Consequently, both polypeptides were dissolved in TFE prior to wet-spinning in acetone, whilst the polypeptide concentration was increased from 15 to 30 %wt./vol., in order to minimise the formation of internal pores and ensure mechanical stability in the resultant wet-spun fibres. As-spun fibres were then reacted with activated Ph or crosslinked via an EDC-mediated condensation reaction and investigated to determine their mechanical tensile properties.

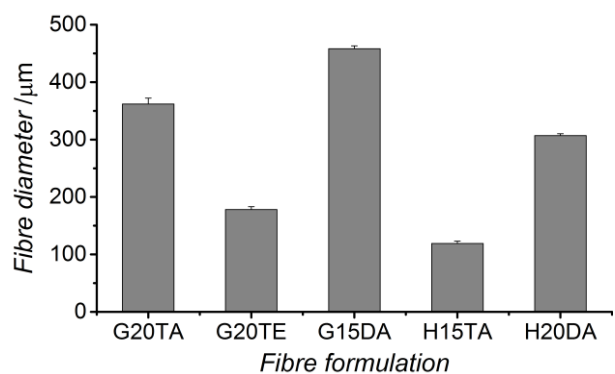

**Fig. 5.** Diameter distribution in fibres wet-spun from polypeptide solutions prepared with varied solution concentrations, wet-spinning solvents and coagulating media.

### 3.3 Tensile and swelling properties of crosslinked wet-spun fibres

HFC and gelatin polypeptides were reacted with activated Ph in the wet-spun fibre state, in order to preserve fibre morphology in hydrated conditions via the formation of a covalent network at the molecular level. A carbodiimide-mediated crosslinking reaction was also carried out as a state-of-the-art crosslinking method. ATR-FTIR on reacted fibres was carried out to investigate the potential detection of bands associated with crosslinking-related amide bonds (Figure 3, (c) and (d)). Resulting spectra confirmed the chemical compositions of native polypeptides, although no crosslink amide bonds could be identified, likely due to the overlap of these with polypeptide bands of amide bonds. Other than ATR-FTIR, quantitative confirmation of covalent crosslinking was obtained via TNBS assay (Table 2). Reacted fibres displayed a degree of crosslinking of at least 40 mol.-% (with regard to the polypeptide lysine content), with Ph-crosslinked samples showing the highest $C$ values ($C$: 51±1 mol.-%). The formation of a covalent network at the molecular level is likely to explain the wet stability of reacted fibres in physiological conditions (PBS, 37 °C), in contrast to the instantaneous dissolution of as-spun samples upon contact with the aqueous medium. Samples G30TA-Ph displayed nearly-complete degradation ($M_L$: 99±1wt.-%) following 5-day incubation in PBS, reflecting the fact that only about 50 mol.-% of collagen lysines were functionalised following reaction with NHS-activated Ph. An increase in network crosslink density as well as the employment of polypeptides with increased molecular weight would be expected to promote the formation of wet-spun fibres with delayed degradation profiles, as observed in the case of synthetic polyester urethane networks [34].

Dry tensile testing could be successfully carried out in all the reacted samples, which was not the case with some of the as-spun fibres. In carbodiimide-crosslinked samples, an elasto-plastic deformation was observed following application of mechanical loads, whilst the stress-strain curves of Ph-crosslinked fibres revealed poorly defined yield points (Fig. 6, A). The formulation-dependent mechanical response can be explained by considering the variation in molecular network architecture in crosslinked fibres as well as in light of the glassy state of dry gelatin at room temperature [3, 5, 7, 21, 24]. Reaction with activated Ph mainly leads to the formation of intermolecular crosslinks, resulting in increased elasticity in comparison to carbodiimide-crosslinked networks, as mainly based on intramolecular crosslinks [35]. Moreover, the fact that no post-drawing of fibres was carried out, leading to polypeptide chains being covalently-crosslinked in an amorphous, glassy state [3, 5, 7, 21], may also account for the observed tensile behaviour.

Variation in young's moduli, maximal tensile strength and elongation at break were correlated with changes in degree of crosslinking across sample formulations (Table 2). The highest tensile modulus ($E$: 844 ± 85 kPa) was measured in gelatin samples crosslinked with activated Ph, whilst $E$ varied in the range of 588 ± 38 – 726 ± 43 kPa in gelatin samples crosslinked with either EDC/NHS-mediated condensation reaction or activated Ph (in the presence of activated EDC-NHS), respectively. Likewise, samples G30TA-Ph displayed the highest elongation at break ($\varepsilon_b$ ~ 11%), whilst samples G30TA-Ph$^*$ and G30TA-EN displayed a somewhat lower value ($\varepsilon_b$ ~ 9%). Overall, observed values of $E$ and $\sigma_{max}$ appeared to be somewhat lower than the ones of wet-spun post-drawn gelatin fibres [3, 4, 5], in agreement with the random polypeptide organisation and the presence of pores in the as-spun fibre [15]. When moving from gelatin to HFC samples, a more than three orders of magnitude reduction in tensile modulus was observed, whilst nearly 4% elongation at break was measured (Table 2; Fig. 6, B). These results reflect the variation in molecular weight between gelatin and HFC (Fig. 2, B), which is in agreement with previous observations on solution viscosity (Fig. 2, A) and wet-spun morphology (Fig. 6).

The increase in crosslink density and corresponding tensile moduli in samples G30TA-Ph suggest that the diacid crosslinking route leads to an increased yield of crosslink density in comparison to either the state-of-the-art carbodiimide-mediated condensation reaction or the Ph functionalisation reaction in the presence of EDC/NHS. In samples Ph, intermolecular crosslinks are likely to be to be formed following nucleophilic addition of amine terminations of the polypeptide with activated Ph terminations. In contrast, carbodiimide-mediated condensation reaction between polypeptide carboxylic acid and amino terminations can only result in zero-length intramolecular crosslinks, so that only slightly modified mechanical properties can be accomplished [26, 35]. Interestingly, the deactivation of EDC/NHS following activation of Ph and prior to the crosslinking reaction with gelatin resulted in a slight increase in crosslink density in samples G30TA-Ph in comparison to samples G30TA-Ph[*] (Table 2).

**Table 2.** Young's modulus ($E$), maximal tensile stress ($\sigma_{max}$), elongation at break ($\varepsilon_b$), degree of crosslinking ($C$) and equilibrium swelling ratio (PBS, pH 7.4) of covalently crosslinked fibres. [a] Exemplary tensile properties of PBS-equilibrated crosslinked fibres.

| Sample ID | $E$ /MPa | $\sigma_{max}$ /MPa | $\varepsilon_b$ /% | SR /wt.-% | $C$ /mol.-% |
|---|---|---|---|---|---|
| G30TA-Ph[*] | 726 ± 43 | 65 ± 12 | 9 ± 1 | 246 ± 33 | 40 ± 1 |
| G30TA-Ph | 844 ± 85 | 52 ± 3 | 11 ± 0 | 382 ± 31 | 51 ± 1 |
| G30TA-EN | 588 ± 38 (0.23 [a]) | 31 ± 3 (0.093 [a]) | 9 ± 0 (39 [a]) | 296 ± 37 | 41 ± 1 |
| HFC30TA-EN | 0.5 | 1.6 | 4 | n.d. | n.d. |

This observation may suggest that the EDC/NHS-mediated condensation reaction is likely to act as a competing reaction during Ph-mediated crosslinking. However, the increased tensile modulus in samples G30TA-Ph[*] compared to samples G30TA-EN gives evidence that the Ph-mediated crosslinking reaction occurs preferentially in comparison to carbodiimide-mediated condensation, given that a comparable degree of crosslinking was observed in previously-mentioned samples (Table 2).

Other than the dry state, PBS-equilibrated wet-spun samples displayed a swelling ratio in the range of 246 ± 33 – 382 ± 31 wt.-%, with samples Ph taking up more PBS than samples Ph[*] and EN. This is somewhat counterintuitive given that the latter samples exhibited increased crosslink density and tensile modulus. The most logical explanation for this observation is that the reaction with activated Ph molecules promotes a small yield of polypeptide grafting concomitant to the formation of a covalent network, due to the high excess of Ph in the reacting mixture. Grafted carboxylic terminations of Ph are likely to mediate hydrogen bonding, thereby promoting a higher PBS uptake in comparison with samples Ph[*] and EN.

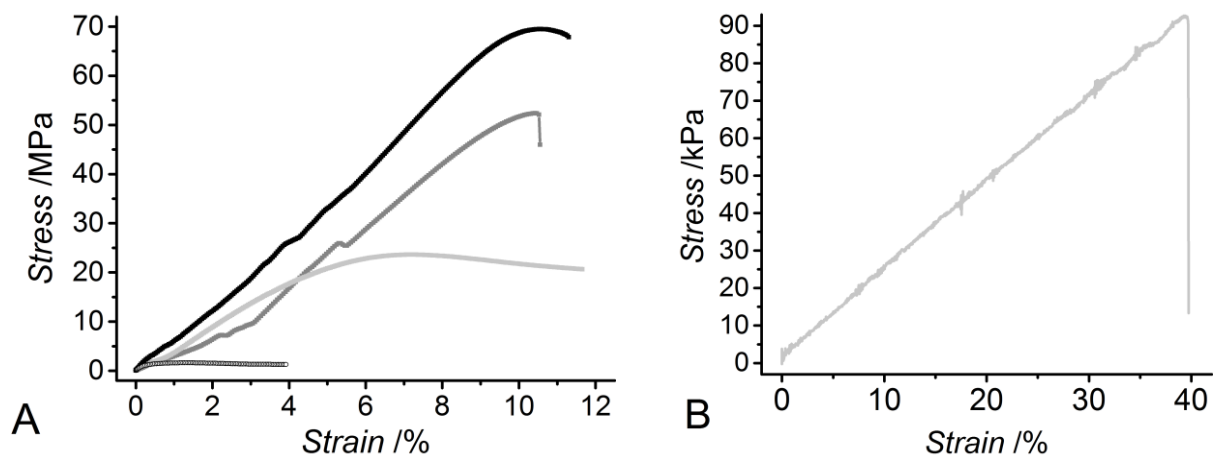

**Fig. 6.** (A): exemplary stress-strain tensile curves of dry covalently-crosslinked fibres. (–■–): G30TA-Ph[*], (–■–): G30TA-Ph, (–■–): G30TA-EN, (–○–): H30TA-EN. (B): example of stress-strain tensile curve (–■–) of PBS-equilibrated covalently-crosslinked sample G30TA-EN.

Other than samples G30TA-Ph, samples G30TA-Ph[*] showed a lower swelling ratio compared to G30TA-EN, despite the comparable degree of crosslinking between the two systems (Table 2). This observation provides further evidence of the preferential occurrence of Ph-mediated crosslinking with respect to carbodiimide-mediated crosslinking reaction. The presence of Ph aromatic segments in resulting covalent network is likely to promote the formation of pi-pi aromatic interactions between crosslinked polypeptide chains [19], thereby accounting for the lower *SR* values in samples Ph[*] compared to EN.

Following PBS equilibration of crosslinked fibres, an exemplarily tensile test was carried out on sample G30TA-EN, in order to elucidate the mechanical behaviour of hydrated samples. Resulting tensile behaviour was described by a linear stress-strain curve, whereby the sample could be stretched up to 40% of their initial fibre length, corresponding to a four-time increase of $\varepsilon_b$ with respect to the dry samples. Other than $\varepsilon_b$, the Young's modulus (*E*) was dropped from 588 MPa to about 230 kPa, whilst the maximal stress ($\sigma_{max}$) was reduced from 31 MPa to 93 kPa (Table 2). Such stark variation in mechanical properties from the dry to the wet state have been observed in other crosslinked gelatin systems and found to be associated with a water-induced plasticising effect. In hydrated conditions, the glass transition temperature of gelatin is decreased from 45 to 20 °C, resulting in a polypeptide morphology transition from a glassy (dry) to a rubbery-like (wet) state [24]. The water-induced increase of chain mobility in PBS-equilibrated crosslinked fibres macroscopically leads to an increase in $\varepsilon_b$ and concomitant decrease in $E$ and $\sigma_{max}$.

### 3.4 Crosslinked fibre morphology

Following characterisation of the dry and hydrated fibre properties, the crosslinked fibre morphology was also addressed. Fig. 7 displays SEM images of fibres wet-spun from TFE solutions (30 %wt./vol. polypeptide) before and after the crosslinking reaction (with either Ph$^*$, Ph or EN). As expected, the increase in polypeptide concentration in the wet-spinning solution (15 → 30 %wt./vol.) led to gelatin-based wet-spun fibres with a dense internal morphology (Fig. 6, A), whilst small voids ($P$: 18±8 μm) were still observed in the HFC-based fibres (Fig. 7, E). Interestingly, the increase in HFC concentration in the wet-spinning solution resulted in fibres with an increased pore size. The most logical explanation for this finding is that pore coalescence occurred during the wet-spinning phase separation in light of the increase in polypeptide solution concentration and corresponding increase in fibre density. It is therefore expected that a further increase in HFC concentration in the wet-spinning solution may suppress the presence of internal pores in resulting fibres. Following the crosslinking reaction, gelatin samples displayed a nearly-homogeneous fibre surface and cross-section (Fig. 7, B-H), whilst small surface defects were observed in retrieved low-molecular weight samples (Fig. 7, F-H), regardless of the crosslinking method applied. The latter topological features are likely to be attributable to the rapid drying of the reacted sample following evaporation of ethanol. Together with morphological observations, crosslinked fibre diameters were also quantified, whereby a uniform diameter was observed in HFC-based (*D*: 212 ± 6 – 243 ± 2 μm) compared to gelatin-based (*D*: 341 ± 8 – 355 ± 28 μm) samples (Table 3).

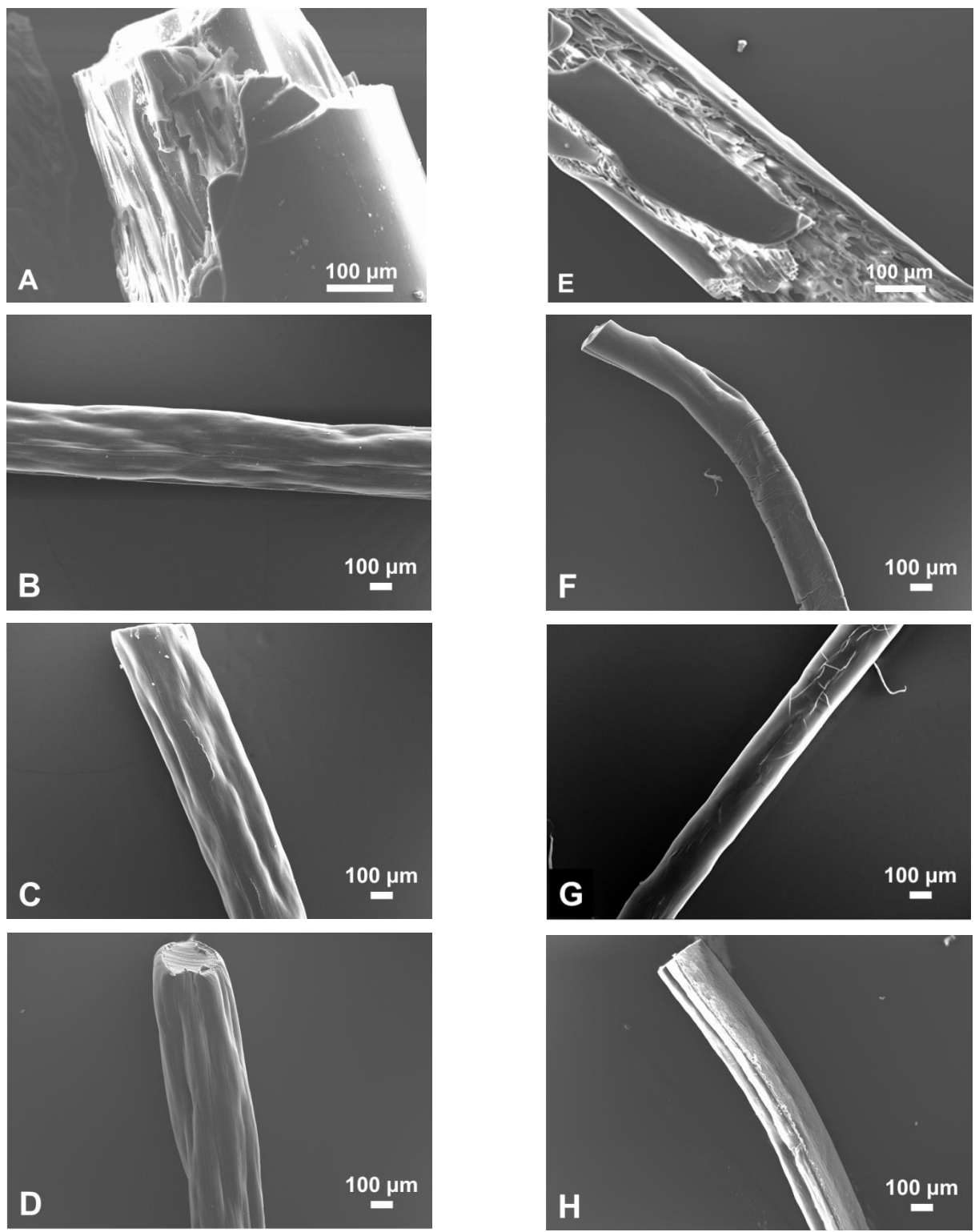

**Fig. 7.** SEM of gelatin (A-D) and HFC (E-H) fibres following either wet-spinning or after the crosslinking reaction. (A): G30TA; (B): G30TA-Ph*; (C): G30TA-Ph; (D): G30TA-EN; (E): H30TA; (F): H30TA-Ph*; (G): H30TA-Ph; (H): H30TA-EN.

These observations therefore suggest that the crosslinking reaction had a minimal impact on the fibre morphology of wet-spun samples and that wet-spinning parameters, particularly the solvent properties, coagulating medium and polypeptide molecular weight have a greater influence.

**Table 3.** Fibre diameters ($D$) of gelatin and HFC-based samples following reaction with either Ph[*], Ph or EN.

| Sample ID | G30TA-Ph[*] | G30TA-Ph | G30TA-EN | H30TA-Ph[*] | H30TA-Ph | H30TA-EN |
|---|---|---|---|---|---|---|
| $D$ /μm | 352 ± 6 | 341 ± 8 | 355 ± 28 | 212 ± 6 | 243 ± 2 | 238 ± 18 |

## 4. Conclusions

Widely available collagen-derived polypeptides of varied molecular weight were evaluated for the preparation of mechanically stable wet-spun biomimetic fibres. Both sample polypeptides were wet-spun against common non-solvents for collagen, such as acetone and ethanol, and both lead to fibre formation. Relatively smooth fibre surfaces were obtained, whilst the internal fibre morphology was strongly affected by the wet-spinning solvent viscosity and polarity, so that a spectrum of porous geometries was observed, depending on spinning conditions and material properties. HFC as low molecular ($M_n$: 17 kDa) polypeptide mostly resulted in porous fibres ($P$: 4±1 μm) when the same polypeptide concentration as in the case of gelatin (as high molecular weight backbone, $M_n$: 106 kDa) was applied, although an inverse correlation was apparent between polypeptide concentration and pore formation. Dry-state tensile tests were successfully carried out on crosslinked fibres, revealing that tensile moduli ($E$: 0.5 – 844 MPa) and elongation at break ($\varepsilon_b$: 4 – 9 %) could be adjusted based on the type of covalent network at the molecular level and polypeptide building block used during wet-spinning. Fibres retained their morphology in PBS, following polypeptide functionalisation with either NHS-activated Ph or carbodiimide-mediated condensation reaction in the wet-spun state. In light of the observed wet-spinnability, both polypeptides could be applied as single or combined building blocks for the design of biomimetic fabrics for applications in chronic wound care or as tissue barriers for dental tissue repair.

**Acknowledgements**

The authors wish to acknowledge support from The Clothworkers' Centre for Textile Materials Innovation for Healthcare and the Clothworkers' Foundation. The authors would like to thank J. Hudson as well as M. Fuller and A. Hewitt for their help with SEM and mechanical testing, respectively. N. Pechlivani and S. Brookes as well as L. Johnson are greatly acknowledged for their assistance with SDS-page and ATR-FTIR spectroscopy, respectively.